\documentclass[letterpaper, 10 pt, conference]{ieeeconf}  

\IEEEoverridecommandlockouts                              

\usepackage{times} 
\usepackage{authblk}
\usepackage{float}
\usepackage{graphicx}
\usepackage{amsmath}
\usepackage{multirow}
\usepackage{booktabs}
\usepackage{hyperref}
\usepackage{eso-pic}
 
\title{\LARGE \bf
CM-UNet: A Self-Supervised Learning-Based Model for Coronary Artery Segmentation in X-Ray Angiography}

\author[1]{Camille Challier}
\author[1]{Xiaowu Sun}
\author[3]{Thabo Mahendiran}
\author[1]{Ortal Senouf}
\author[2]{Bernard De Bruyne}
\author[3]{Denise Auberson}
\author[3]{\\Olivier Müller}
\author[3]{Stephane Fournier}
\author[1]{Pascal Frossard}
\author[1]{Emmanuel Abbé}
\author[1]{Dorina Thanou}

\affil[1]{EPFL, Switzerland}
\affil[2]{Cardiovascular Center OLV, Belgium}
\affil[3]{CHUV, Switzerland}

\begin{document}

\maketitle

\AddToShipoutPictureBG*{%
  \AtTextLowerLeft{%
    \put(0,-30){%
      \parbox{\textwidth}{%
        \scriptsize \centering
        ©2025 IEEE. Personal use of this material is permitted. Permission from IEEE must be obtained for all other uses, in any current or future media, including reprinting/republishing this material for advertising or promotional purposes, creating new collective works, for resale or redistribution to servers or lists, or reuse of any copyrighted component of this work in other works.
      }
    }%
  }%
}

\thispagestyle{empty}
\pagestyle{empty}

\begin{abstract}
Accurate segmentation of coronary arteries remains a significant challenge in clinical practice, hindering the ability to effectively diagnose and manage coronary artery disease. The lack of large, annotated datasets for model training exacerbates this issue, limiting the development of automated tools that could assist radiologists. To address this, we introduce CM-UNet, which leverages self-supervised pre-training on unannotated datasets and transfer learning on limited annotated data, enabling accurate disease detection while minimizing the need for extensive manual annotations. Fine-tuning CM-UNet with only 18 annotated images instead of 500 resulted in a 15.2\% decrease in Dice score, compared to a 46.5\% drop in baseline models without pre-training. This demonstrates that self-supervised learning can enhance segmentation performance and reduce dependence on large datasets. This is one of the first studies to highlight the importance of self-supervised learning in improving coronary artery segmentation from X-ray angiography, with potential implications for advancing diagnostic accuracy in clinical practice. The source code is publicly available at 
https://github.com/CamilleChallier/Contrastive-Masked-UNet.

{\textbf{\textit{Clinical Relevance}}}\textemdash By enhancing segmentation accuracy in X-ray angiography images, the proposed approach aims to improve clinical workflows, reduce radiologists' workload, and accelerate disease detection, ultimately contributing to better patient outcomes. 

\end{abstract}

\section{INTRODUCTION}

Coronary artery disease (CAD) is one of the leading causes of mortality worldwide, affecting millions of individuals. CAD is defined by the accumulation of plaque in coronary arteries, leading to vessel narrowing (stenosis), which restricts blood circulation and may lead to serious cardiovascular complications. Invasive X-ray coronary angiography (ICA) is the gold standard for CAD diagnosis, where contrast agents are injected to visualize the arteries through X-ray projections from multiple angles. Clinicians typically assess stenosis severity by visually inspecting the thickness of arteries in angiographic images, a process that is time-intensive and relies on subjective expertise. 

Accurate segmentation of coronary arteries is essential for reliable stenosis quantification, enabling precise diagnosis and treatment for CAD. Automated segmentation methods offer significant advantages over traditional manual segmentation by providing objective and precise quantification analysis. However, coronary artery segmentation still presents many challenges. First, the lack of sufficiently large, annotated, and publicly available datasets limits deep learning model training \cite{popov_dataset_2024}. Additionally, the coronary arteries are extremely fine-grained structures, often approaching the resolution limits of X-ray imaging systems and complicating precise visualization and analysis \cite{ghekiere_image_2017}. The highly variable topology of coronary artery trees makes it difficult to incorporate prior shape knowledge effectively. Furthermore, the low signal-to-noise ratio in coronary imaging, combined with overlapping structures such as the catheter, spine, and rib cage, further impedes accurate segmentation \cite{lin_extraction_2005}.

To address these challenges, we propose a self-supervised learning (SSL) based method for coronary artery segmentation from ICA imaging using limited labeled data. SSL offers a promising solution to overcome the lack of large annotated datasets needed for training in the medical imaging domain. It also addresses the issue of variations in image quality and settings across different medical labs, which often hinder the sharing of datasets between institutions. SSL employs vast amounts of unlabeled data during pre-training, allowing models to develop robust image representations while reducing the need for labeled datasets \cite{huang_self-supervised_2023}.
Our novel segmentation framework, namely Contrastive Masked UNet (CM-UNet) incorporates Contrastive Masked Auto-Encoder (CMAE) \cite{huang_contrastive_2024} with a UNet backbone. Furthermore, we conduct a comprehensive evaluation and benchmarking of state-of-the-art (SOTA) SSL methods using a UNet architecture, specifically applied for coronary artery segmentation. Our experiments demonstrate that CM-Unet achieves substantial performance improvements over other SSL-based methods, particularly in low-data settings when applied to the FAME2 (Fractional Flow Reserve Versus Angiography for Multivessel Evaluation 2) dataset \cite{de_bruyne_fractional_2012}.
These advancements highlight CM-Unet's potential to enhance coronary artery segmentation accuracy, ultimately aiding in the diagnosis and treatment of cardiovascular diseases. The reduced need for annotated data could enable broader clinical application, ultimately contributing to more accessible cardiovascular care.

\section{Related Works}
\label{sec:related_work}
\subsection{Supervised learning for ICA images analysis}
CNN-based architectures have demonstrated remarkable performance in medical image segmentation tasks \cite{krizhevsky_imagenet_2012}. For ICA, deep learning models like UNet \cite{ronneberger_u-net_2015}, ResNet \cite{he_deep_2015}, and DenseNet \cite{huang_densely_2018} can effectively segment major vessels \cite{yang_deep_2019}. Building upon this foundation, AngioNet \cite{iyer_angionet_2021} addresses specific challenges such as poor contrast and unclear vessel boundaries, improving segmentation accuracy. AngioPy \cite{mahendiran_angiopy_2025}, a deep learning model for coronary segmentation, integrates user-defined ground-truth points to enhance performance and reduce the need for manual correction.

However, supervised models typically learn task-specific features, making them less adaptable across different datasets and clinical applications. The dependency on labeled data also presents challenges in ICA segmentation, as manual annotation is time-consuming and error-prone due to overlapping structures, low-contrast images, and the fine-grained nature of coronary arteries. These limitations highlight the need for approaches that can leverage large amounts of unlabeled data to learn robust and generalized representations. 

\subsection{Self-supervised learning for medical images analysis}
SSL offers a promising alternative by enabling models to learn effective representations from unlabeled data, reducing the reliance on manual annotations. It has been explored in medical imaging, with notable successes in pre-training models for segmentation and classification tasks. SSL methods can be categorized into four types: Innate Relationship, Generative, Contrastive, and Self-Prediction \cite{huang_self-supervised_2023}. Innate Relationship consists of pre-training a model on a custom-designed task and has been applied to various medical datasets, including echocardiogram imagery \cite{dezaki_echo-rhythm_2021} and ultrasound data \cite{jiao_self-supervised_2020}. Generative SSL frameworks like Models Genesis \cite{zhou_models_2019} are used to pre-train UNet architectures on 3D anatomical structures in CT, X-ray, and MRI images. Contrastive SSL approaches, such as SimCLR \cite{chen_simple_2020} and MoCo \cite{he_momentum_2020}, have been adapted for medical image analysis, including CT classification tasks \cite{wolf_self-supervised_2023} and volumetric image segmentation in MRI \cite{chaitanya_contrastive_2020}. These methods effectively learn meaningful image representations by maximizing agreement between augmented views from the same image. Self-prediction methods, such as masked auto-encoders (MAE) \cite{he_masked_2021}, have recently gained popularity. Tian et al. proposed Spark \cite{tian_designing_2023}, a sparse convolution-based adaptation of MAE for CNNs, which surpasses contrastive methods. Spark has demonstrated promising results across various CT classification tasks \cite{wolf_self-supervised_2023}.

While numerous SSL methods have been explored in medical imaging, their application to ICA analysis remains underexplored. Ma et al. proposed a self-supervised approach that leverages adversarial learning for ICA vessel segmentation \cite{ma_self-supervised_2021}. Similarly, Zeng et al. proposed a self-supervised method for single-frame subtraction and vessel segmentation in ICA, achieving good performance with minimal fine-tuning on a limited annotated dataset \cite{zeng_pretrained_2024}. However, challenges such as capturing fine-grained vessel structures, variable coronary topology, and low signal-to-noise ratios in ICA images remain largely unresolved.

\section{Self-Supervised Learning-based Framework for Coronary Artery Segmentation}

\begin{figure}[h]
    \centering
    \includegraphics[width = 0.96\columnwidth]{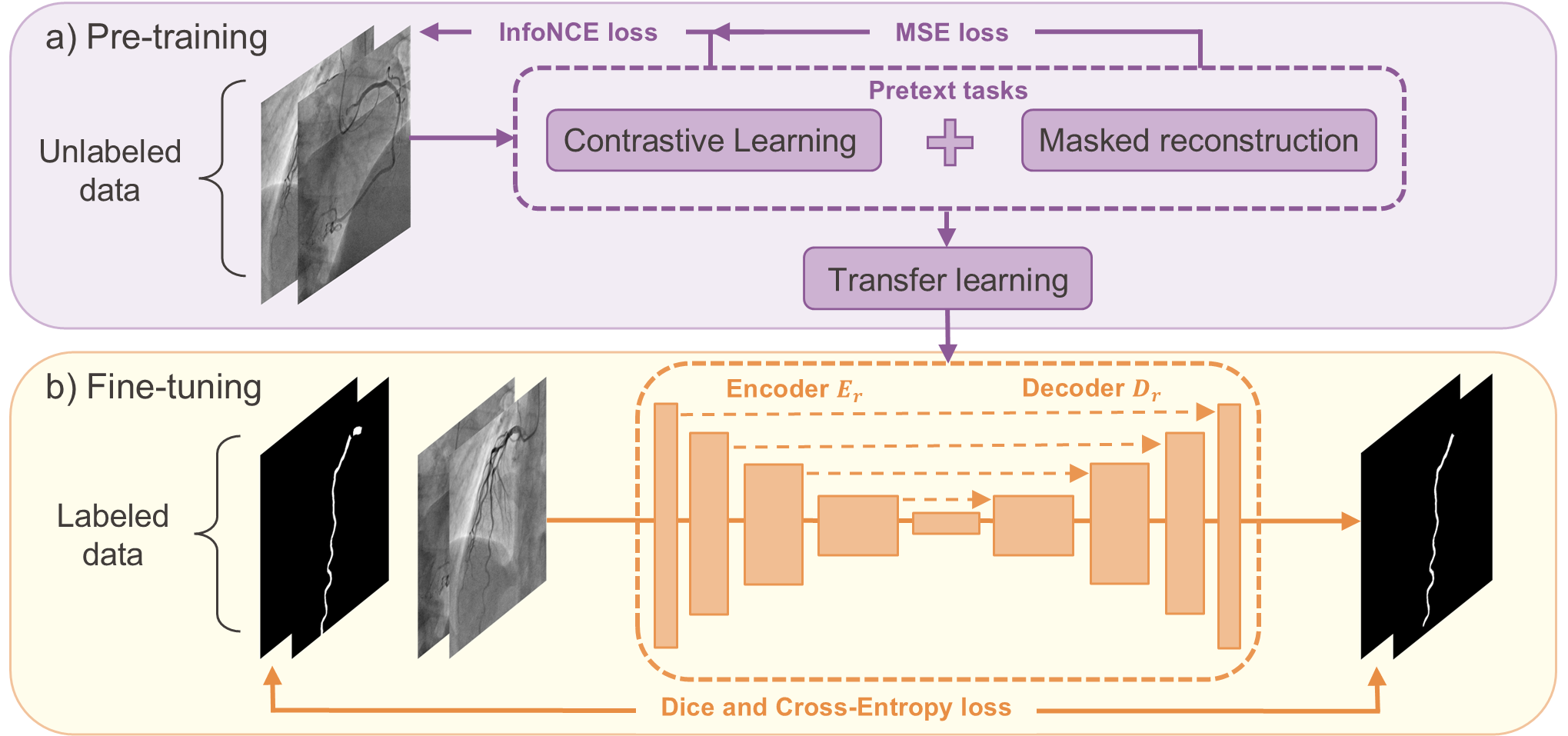}
    \caption{Overview of the CM-UNet training pipeline, including both pre-training and fine-tuning stages. For further details on the pre-training process, refer to Figure \ref{fig:CM-UNet_archi}.}
    \label{fig:training_pipeline}    
\end{figure}

In this section, we introduce CM-UNet, an SSL framework for coronary artery segmentation from ICA images. Our approach builds on the strengths of contrastive learning and MAE to effectively learn both global representations and fine-grained anatomical details. The proposed framework follows a two-step process, as illustrated in Fig. \ref{fig:training_pipeline}. The first step is pre-training, a self-supervised stage that leverages contrastive learning and masked image reconstruction to learn robust feature representations from unlabeled data. The second step is fine-tuning, a supervised refinement phase where the pre-trained model is adapted to the coronary artery segmentation task using a limited number of labeled images.

\begin{figure*}[htbp]
\smallskip
    \centering
    \includegraphics[width = 0.79\paperwidth]{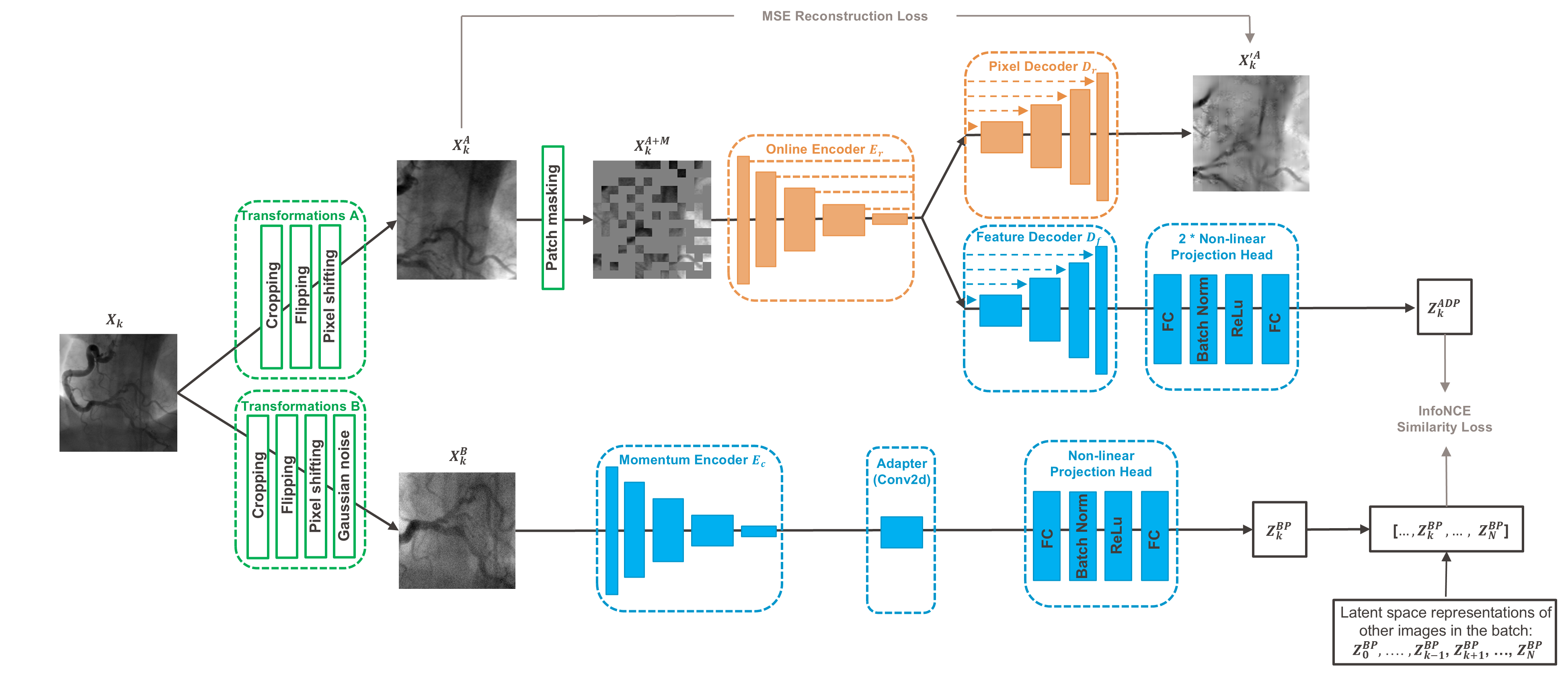}
    \caption{Illustration of Contrastive Masked UNet (CM-UNet) method for self-supervised pre-training. The framework consists of two key components: an online encoder-decoder (in orange) for reconstructing masked image patches, and a momentum branch (in blue) that generates contrastive embeddings. An auxiliary feature decoder is integrated into the online branch to refine latent representations for contrastive learning. The feature decoder and momentum encoder's outputs are compared through a contrastive loss to learn discriminative features. The model is trained using a combination of reconstruction and contrastive loss. The diagram illustrates an example for image k.}
    \label{fig:CM-UNet_archi}    
\end{figure*}

Contrastive learning and MAE represent two of the most prominent approaches, offering complementary strengths in SSL. Contrastive methods excel at learning global representations by distinguishing between different instances, making them effective for tasks requiring robust, instance-level feature discrimination. However, they often struggle with fine-grained structures, especially in small datasets, where instance diversity is limited. On the other hand, MAE methods focus on reconstructing missing image regions, capturing local and contextual information crucial for medical images like ICA, which feature fine-grained anatomical structures. 

By combining the global representation power of contrastive learning with the local reconstruction ability of MAE, we aim to create a more balanced pre-training framework that effectively handles both high-level and fine-grained features in coronary artery segmentation.

\subsection{Self-supervised pre-training with Contrastive Masked UNet}

CM-Unet is composed of two image transformation blocks and two branches, including a reconstruction branch with an encoder $E_r$ and a decoder $D_r$ to reconstruct masked images and a contrastive learning branch to provide contrastive supervision. An illustration of CM-UNet’s framework is depicted in Fig. \ref{fig:CM-UNet_archi}. In the following sections, the three main components of CM-UNet are described: image transformation, which generates diverse augmented views to enhance representation learning; the reconstruction branch, responsible for recovering masked image regions to capture fine-grained anatomical structures; and the contrastive learning branch, which ensures global feature consistency and discriminative representation learning.

\subsubsection{Image transformation}
Given an unlabeled image $X_k$, two random transformations are applied to generate two distinct variations of $X_k$. In the reconstruction branch, transformations including random cropping, flipping, and pixel shifting are applied to produce the augmented image $X^{A}_{k}$. Similarly, in the contrastive branch, transformations such as random cropping, flipping, pixel shifting, and Gaussian noise are used to generate $X^{B}_{k}$. The cropping technique first extracts a base image using resized random cropping. Both branches then generate views by slightly shifting the cropping locations with the pixel-shifting technique, ensuring overlapping regions for positive pairs \cite{huang_contrastive_2024}.
In the reconstruction branch, masking is additionally applied to create the transformed masked image $X^{A+M}_{k}$.

\subsubsection{Reconstruction Branch}
The reconstruction encoder-decoder ($E_r$, $D_r$) employes a UNet as backbone to reconstruct $X^{A}_k$ from $X^{A+M}_{k}$. The online encoder $E_r$ maps the masked image $X^{A+M}_{k}$ to its latent representation denoted as $Z^{A}_k$. Subsequently, the pixel decoder $D_r$ maps the features $Z^{A}_k$ to the reconstructed image $X^{\prime A}_{k}$. Mean Square Error (MSE) is used as the loss function to compute the reconstruction error on the masked patches between $X^{A}_k$ and $X^{\prime A}_{k}$:
\begin{equation}
L_r = \frac{1}{N_m} \sum_{m} (X^{\prime A}_{k_m} - X^{A}_{k_m})^2, 
\end{equation}
with $N_m$ denoting the number of masked patches.
Through this process, $E_r$ and $D_r$ collaboratively learn to reconstruct the masked patches, enabling the model to develop a unified representation for each patch within the image.

\subsubsection{Contrastive learning Branch}
The contrastive learning branch aims to guide the reconstruction encoder $E_r$ in acquiring more distinctive features. Therefore, the encoder $E_c$ in the contrastive learning branch shares the same architecture as the encoder $E_r$. $E_c$ processes the augmented image $X^{B}_k$, transforming it into a feature embedding. This embedding is then passed through an adapter for shape compatibility, implemented as a Conv2D layer, resulting in $Z^{B}_k$.

To align the output of the contrastive encoder and the output of the reconstruction encoder, an auxiliary feature decoder $D_f$, is then applied to transfer the feature embedding $Z^{A}_k$  to $Z^{AD}_k$. While both the feature decoder and pixel decoder share the same architecture, their parameters are not shared, as they serve different learning objectives.

To map these two representations ($Z^{B}_k$ and $Z^{AD}_k$) to the space where contrastive loss is applied, the ``projection-prediction" head is added after the feature decoder and the ``projection" head after the contrastive learning encoder.
The projection head  $g$ is a multi-layer perceptron (MLP) with two hidden layers. 
The architecture consists of a fully connected layer (FC), followed by batch normalization, a ReLU activation, and a second hidden layer.
The predictor $q$ uses the same architecture as $g$. Thus, the following representations are obtained: $Z^{BP}_k = g(Z^{B}_k)$ and $Z^{ADP}_k = q(g(Z^{AD}_k))$.

To ensure that the online encoder's representations capture holistic and discriminative features while improving generalization performance, the InfoNCE loss \cite{oord_representation_2019} is applied for contrastive learning. This loss compares two embeddings: $Z^{BP}_k$, the output of the contrastive learning encoder, and $Z^{ADP}_k$, the output from the feature decoder $D_f$. 

The output $Z^{BP}_k$ is saved alongside the latent space representations of the contrastive learning encoder from other images of the batch, specifically [$Z^{BP}_0$, ..., $Z^{BP}_k-1$, $Z^{BP}_k+1$, ..., $Z^{BP}_N$]. This list thus contains one element, $Z^{BP}_k$, which is derived from the same original image as the output of the reconstruction branch $Z^{ADP}_k$, forming the positive pair, and many elements from different original images, which represent the negative pairs. The encoder model is trained to differentiate between representations from the same original image (positive pairs) and representations that come from different original images (negative pairs).
We compute the cosine similarity $s$ between a pair:
\begin{equation}
s = \frac{Z^{ADP}_m \cdot Z^{BP}_n}{\| Z^{ADP}_m \|_2  \| Z^{BP}_n \|_2},
\end{equation}
where $n$ and $m$ represent two different images.
We note $s_+$ and $s_-$ as the cosine similarities between the positive and negative pairs, respectively.
InfoNCE loss can be written as follows: 
\begin{equation}
L_c = - \log \left( \frac{\exp(s_+ / \tau)}{\exp(s_+ / \tau) + \sum_{j=1}^{K-1} \exp(s_-^j / \tau)} \right),
\label{eq:contrastive_loss}
\end{equation}
where \( \tau \) is a temperature hyperparameter.

The total learning objective of CM-Unet for pre-training is a weighted sum of the reconstruction loss \( L_r \) and the contrastive loss \( L_c \), defined as:
$L = L_r + \lambda_c L_c$.

\subsubsection{Pre-training}
During pre-training, the contrastive learning encoder and projection head parameters are updated using an exponential moving average algorithm (EMA) \cite{huang_contrastive_2024}. 
The parameters of the momentum encoder are updated as a weighted average of the current momentum encoder and online encoder parameters.

\subsection{Downstream task: ICA image segmentation}

Following the pre-training phase, the online reconstruction encoder $E_r$ and the pixel decoder $D_r$ are leveraged for downstream segmentation tasks.
The encoder and decoder are initialized with pre-trained weights and combined to form a UNet model with skip connections. However, the weights and biases of the pre-trained final convolution layer are excluded to enable task-specific adjustments. The UNet model is fine-tuned on the labeled dataset.
A hybrid loss function is used, combining Dice and Cross-Entropy loss:
\begin{equation}
\mathcal{L} =  \mathcal{L}_{\text{Dice}} + \lambda_{CE} \mathcal{L}_{\text{CE}}.
\end{equation}
Dice loss ensures overlap with the ground truth, while Cross-Entropy penalizes pixel-wise mis-classifications. 

\section{Experiments}
\label{sec:results}

\subsection{Dataset} 

The Fractional Flow Reserve Versus Angiography in Multivessel Evaluation 2 RCT (FAME2) dataset \cite{de_bruyne_fractional_2012} used in this work includes 563 patients with stable coronary artery diseases, collected from 28 different clinical sites. The dataset comprises 1,738 labeled ICA images. All studies were approved by the local medical ethics committee, and participants provided written informed consent. Coronary arteries were manually annotated by the cardiology team at Lausanne University Hospital (CHUV).

The data pre-processing pipeline for the FAME2 dataset includes several key steps to enhance image quality and ensure consistency. First, mask smoothing is applied to reduce noise and improve the accuracy of boundary delineation in segmentation masks. The images are then cropped to a fixed size, and corner inpainting is performed to address missing or corrupted pixels. To further refine image quality, unsharp masking filters are used with specific parameters (radius=60, amount=1) to highlight important features. Z-score intensity normalization is applied using the mean and standard deviation of the dataset.
The dataset is divided into training and testing sets, consisting of 1390 and 348 images, respectively. The training set is subsequently split into two subsets: a pre-training set and a fine-tuning set.

\subsection{Evaluation metrics} 


\subsubsection{Segmentation metrics}

\textbf{Dice} measures the overlap between the prediction and ground truth (GT). Equivalent to the F1 score in binary segmentation, it is defined as:  
\begin{equation}
    Dice = \frac{2 TP}{2TP + FP + FN}
\end{equation}
where TP, FP, and FN denote true positives, false positives, and false negatives based on prediction–GT overlap.

\textbf{Centerline Dice (Cl-Dice)} is a Dice coefficient based on the Centerlines of the predicted and GT segmentations \cite{cldice2021}.

\textbf{Intersection over Union (IoU)} also evaluates overlap:  
\begin{equation}
\text{IoU} = \frac{TP}{TP + FP + FN}
\end{equation}




\textbf{Hausdorff Distance (HD)} is the distance between the points on the predicted and GT boundaries, indicating the maximum deviation. Here, the modified HD is used, as it has demonstrated superior performance compared to the directed HD \cite{dubuisson_modified_1994}.
The distance between two points $g$ and $p$ is defined as the Euclidean distance: $d(g, p) = \|g - p\|$. The distance between a point \( g \) and a set of points \( \mathcal{P} = \{p_1, p_2, \dots, p_N\} \) is defined as: $d(g, \mathcal{P}) = \min_{p \in \mathcal{P}} \|g - p\|$.
The directed distance
between two point sets $\mathcal{G} = \{g_1, . . . , g_{N_g\}}$ and point
set $\mathcal{P} = \{ p_1, . . . , p_{N_p\} }$ is given by $d(\mathcal{G}, \mathcal{P}) = \frac{1}{N_g} \sum_{g \in \mathcal{G}} d(g, \mathcal{P})$.
The undirect distance measure of the modified HD is finally denoted by :
\begin{equation}
    HD= max(d(\mathcal{G}, \mathcal{P}), d(\mathcal{P}, \mathcal{G})).
\end{equation}

\subsubsection{Clinical metrics}

\textbf{Arteries Diameter Difference (ADD)} is the difference in diameter between the GT and the predicted coronary arteries. It is computed by first extracting the skeleton of the binary mask and identifying the contour points. The radius at each skeleton point is then calculated by finding the nearest contour point. The final metric is derived by evaluating the mean diameters of the skeleton. 

\subsubsection{Statistical Analysis}
Pearson correlation coefficient (PCC) is introduced to quantify the correlation of the clinical metrics between the manual and automatic segmentation approaches. Bland-Altman analysis \cite{bland_statistical_1986} is used to assess the agreement between the predicted and GT arteries' diameters. 

\subsection{Experimental settings} 

CM-UNet pre-training uses a batch size of 256 and 300 epochs. The learning rate is dynamically adjusted based on the batch size, update interval, and GPU configuration. A linear warmup learning rate schedule is applied for the initial 40 epochs, followed by cosine annealing until epoch 300. Optimization is performed using the AdamW optimizer with betas set to (0.9, 0.95), a weight decay of 0.05, and parameter-specific decay adjustments. The training process incorporated momentum updates, custom hooks, and a fixed random seed to ensure reproducibility. Masking is applied with a patch size of $16\times 16$ and a coverage ratio of 65\%. 

Concerning fine-tuning for the segmentation task, data augmentation techniques such as Gaussian noise, Gaussian blur, random cropping, brightness modification, flipping, and rotations are applied to the fine-tuning set of the FAME2 dataset to enhance model robustness. Three-fold cross-validation is performed on the fine-tuning set to identify the best hyperparameters. Hyperparameter tuning includes a learning rate range from $1 \times 10^{-1}$ to $1 \times 10^{-5}$, batch sizes of 16 and 32, and training over 256 epochs.

All the models are implemented in Pytorch and trained with an NVIDIA A100 GPU with 80 GB memory. 

We compare CM-UNet with several SSL models including
Model Genesis \cite{zhou_models_2019}, MAE \cite{he_masked_2021}, SparK \cite{tian_designing_2023} and MoCoV2  \cite{he_momentum_2020}. All these model employs UNet as the backbone and are evaluated for transfer learning in coronary artery segmentation.
The UNet used in all experiments has four downsampling blocks with 64, 128, 256, and 512 channels, followed by a 1024-channel bottleneck layer. This is paired with a mirrored upsampling path and skip connections. 

\subsection{Model Performance} 

In this work, 20\% of the dataset is reserved for testing segmentation performance across SSL models, while the remaining 80\% is split into pre-training and fine-tuning sets in varying ratios.
The performance of the models is evaluated based on segmentation accuracy, quantified using the Dice score and HD, as well as the PCC to evaluate the relationship between the predicted and GT artery diameters.

\subsubsection{Segmentation results}
\begin{figure}[h]
\smallskip
    \centering
    \includegraphics[width = 0.97\columnwidth]{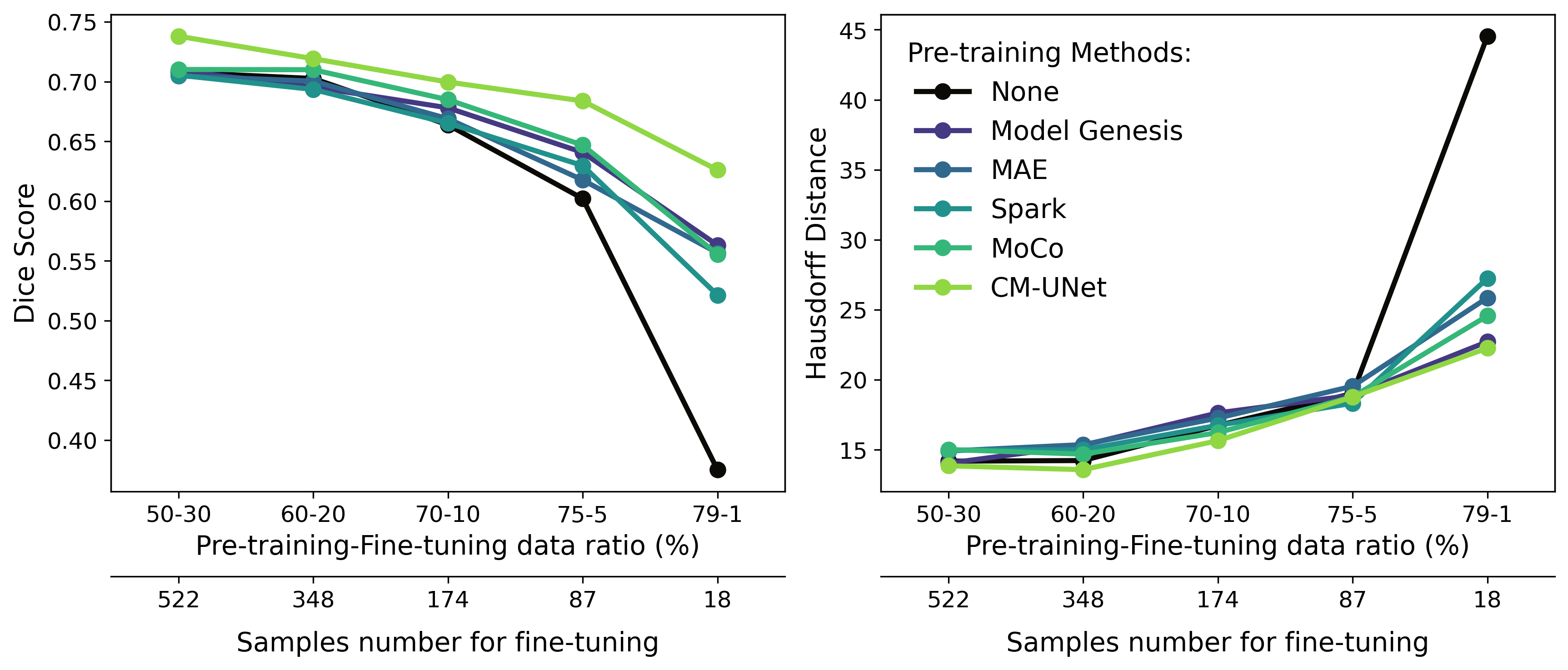}
    \caption{Evalution of the Dice score and Hausdorff distance across various PT-FT ratios and self-supervised methods.}
    \label{fig:size_ss}    
\end{figure}

As shown in Fig. \ref{fig:size_ss}, the pre-trained models consistently outperformed the non-pre-trained model, particularly when less than 10\% of the data (i.e., fewer than 200 images) are available. The performance gap between pre-trained and non-pre-trained models increases as the amount of fine-tuning data decreases, demonstrating the effectiveness of SSL in addressing data-scarce scenarios.
 
For the 70-10 pre-training-fine-tuning (PT-FT) ratio, the mean Dice score of pre-trained models is 0.68, compared to 0.66 for the non-pre-trained model, representing an improvement of 3.03\%. The performance gain across pre-trained models ranges from 0.8\% to 5.4\%. In the case of the 79-1 ratio, where only 1\% of the dataset is used for fine-tuning, pre-trained models achieve a mean Dice score of 0.56, while the non-pre-trained model scored significantly lower at 0.38. This corresponds to a substantial improvement of 49.2\%, with individual pre-trained models showing gains ranging from 38.9\% to 66.8\%. These results highlight the robustness of pre-trained models in scenarios with extremely limited labeled data.
Moreover, when transitioning from the 50-30 to the 79-1 PT-FT ratio: the dice score decreases by 46.5\% (from 0.71 to 0.38) for the non-pre-trained model, and by 22.2\% (0.72 to 0.56) for the pre-trained ones on average.

Among the self-supervised pre-training techniques, CM-UNet outperforms the other methods. For the 50\% pre-training and 30\% fine-tuning ratio, the non-pre-trained model achieves a Dice score of 0.71, while the pre-trained CM-UNet achieves a Dice score of 0.74. This trend is consistent across all PT-FT ratios, further emphasizing CM-UNet's effectiveness in leveraging SSL pre-training for coronary artery segmentation tasks.

We also investigate the performance of SSL methods under extreme data scarcity, where only 1\% of the training dataset (\textit{18 images}) is available for fine-tuning. This analysis is critical for assessing the robustness and generalizability of SSL models in low-data scenarios, which are common in medical imaging tasks. As shown in Table~\ref{tab:metrics}, CM-UNet achieves the highest performance among the SSL methods. Specifically, CM-UNet achieves a Dice score of 0.626, a significant improvement over the supervised UNet model's score of 0.375 and surpassing the highest score of 0.563 achieved by other SSL methods. CM-Unet outperforms competing SSL approaches, yielding significant mean percentage improvements in segmentation metrics: +14.58\% in Dice score, +10.52\% in Cl-Dice score, +15.77 \% in IoU score, along with reductions of -10.90\% in HD and -13.33\% in ADD. This superior performance, despite the higher parameter count (121.49M), demonstrates that the increased complexity enhances segmentation accuracy and robustness compared to models with fewer parameters.

\begin{table}[ht]
    \centering
    \caption{Segmentation performance of supervised (SL) UNet and SSL models with a 79:1 PT-FT ratio.}
    \resizebox{\columnwidth}{!}{
    \setlength{\tabcolsep}{3pt}
    \begin{tabular}{l l c c c c c c}
        \toprule
        \textbf{Method} &\textbf{Model} & \textbf{\# Params} & \textbf{Dice}  &\textbf{Cl-Dice} & 
        \textbf{IoU} & \textbf{HD} & \textbf{ADD}\\
        \toprule
        SL&UNet & 31.04 M &0.375 & 0.417  & 0.248 &  44.54 & 3.929  \\
        \midrule
        \multirow{5}{*}{SSL} 
        &Model Genesis & 31.04 M & 0.563 & 0.584 & 0.414 & 22.72 & 0.893 \\
        &MAE & 31.04 M & 0.557 & 0.589 & 0.410 & 25.85 & 1.066\\
        &Spark & 34.18 M & 0.521 & 0.537 & 0.370 &  27.25 & 1.564\\
        &MoCo &  37.70 M & 0.556 & 0.566 & 0.385& 24.56 & 1.117\\
        &\textbf{CM-UNet} & \textbf{121.49 M}& \textbf{0.626} & \textbf{0.628} & \textbf{0.457}  & \textbf{22.26} & \textbf{0.965}\\

        \bottomrule
        \end{tabular}
        }
    \label{tab:metrics}
\end{table}

The results of this study clearly demonstrate that SSL pre-training is essential for improving segmentation performance, particularly when fine-tuning data is limited. By bridging the gap between global and local feature learning, SSL approaches such as CM-UNet provide a robust foundation for coronary artery segmentation, outperforming traditional non-pre-trained models across a variety of settings.

\subsubsection{Statistical results}

\begin{figure}[h]
    \centering
    \includegraphics[width = 0.94\columnwidth]{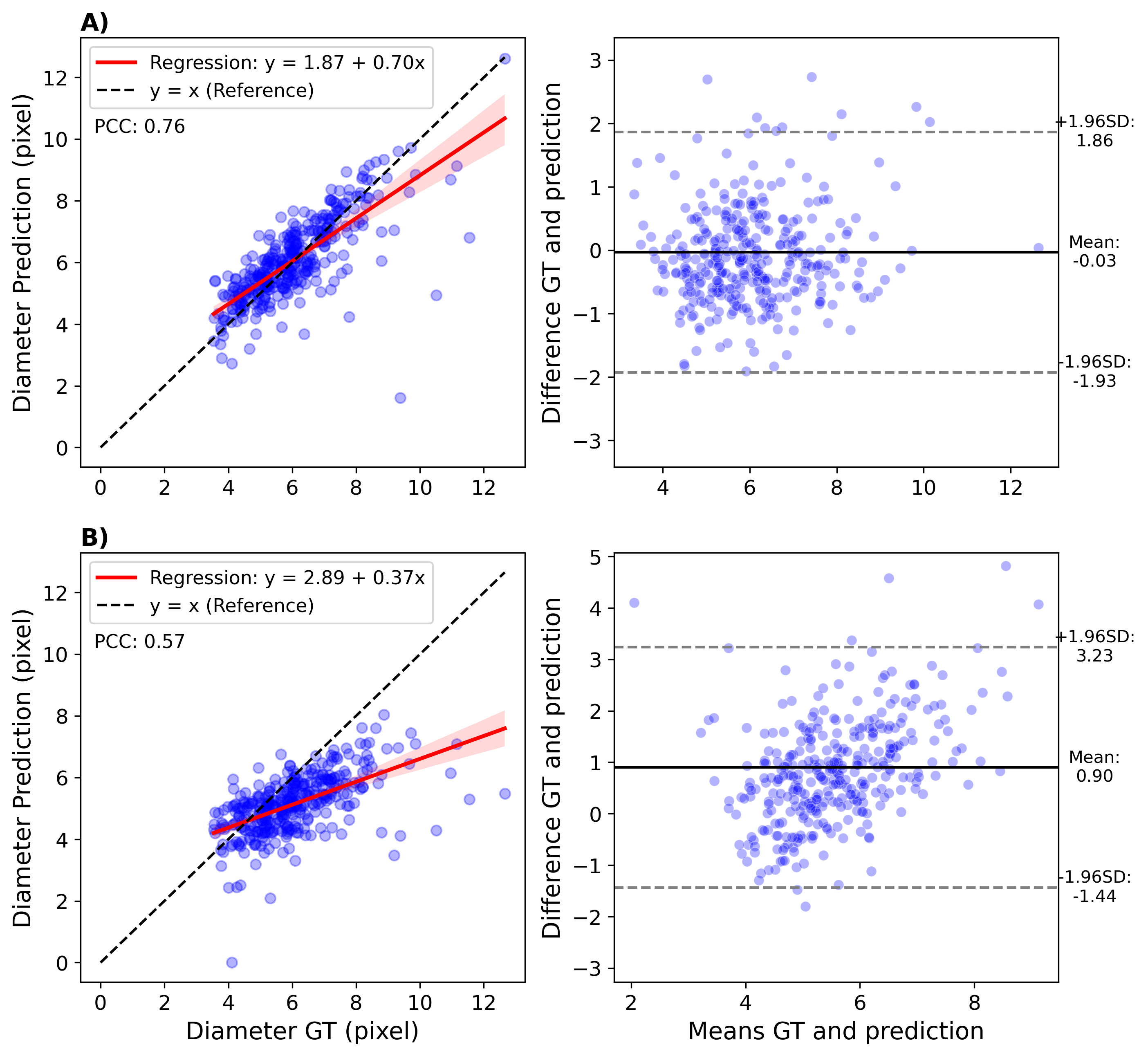}
    \caption{Comparison of CM-UNet diameter predictions with diameter computed on GT annotations. (A) CM-UNet 50:30, (B) CM-UNet 79:1 PT-FT ratio. Left panels: Pearson correlation and linear regression analysis. Right panels: Bland-Altman plots showing agreement across all vessel diameters.}

    \label{fig:pearson}    
\end{figure}

Diameter measurements from CM-UNet with a PT-FT ratio of 50:30 exhibit excellent agreement with GT labels, as shown in Fig. \ref{fig:pearson}. At this ratio, CM-UNet achieves a PCC of 0.76 with p $<$ 0.001 and a mean difference of 0.03 pixels (limits of agreement: -1.93 to 1.86). The p-value, obtained through a t-test, indicates a highly significant linear relationship between the predicted and GT values, with a value of less than 0.001 suggesting strong statistical significance. For the 79:1 ratio, performance is slightly lower but still significant, with a PCC of 0.57 (p $<$ 0.001) and a mean difference of 0.90 pixels (limits of agreement: -1.44 to 3.23). These findings demonstrate CM-UNet's ability to maintain strong agreement with GT labels, even under data-scarce conditions. This highlights the model's robustness and potential to provide clinically acceptable diameter measurements in real-world applications.

\subsubsection{Visualization Results}
This section visualizes and compares CM-UNet's segmentation performance against other models under varying PT-FT ratios. Reconstruction visualizations are also presented to assess how well different SSL approaches capture vessel structures and fine-grained details.

\begin{figure}[h]
    \centering
    \includegraphics[width = 0.96\columnwidth]{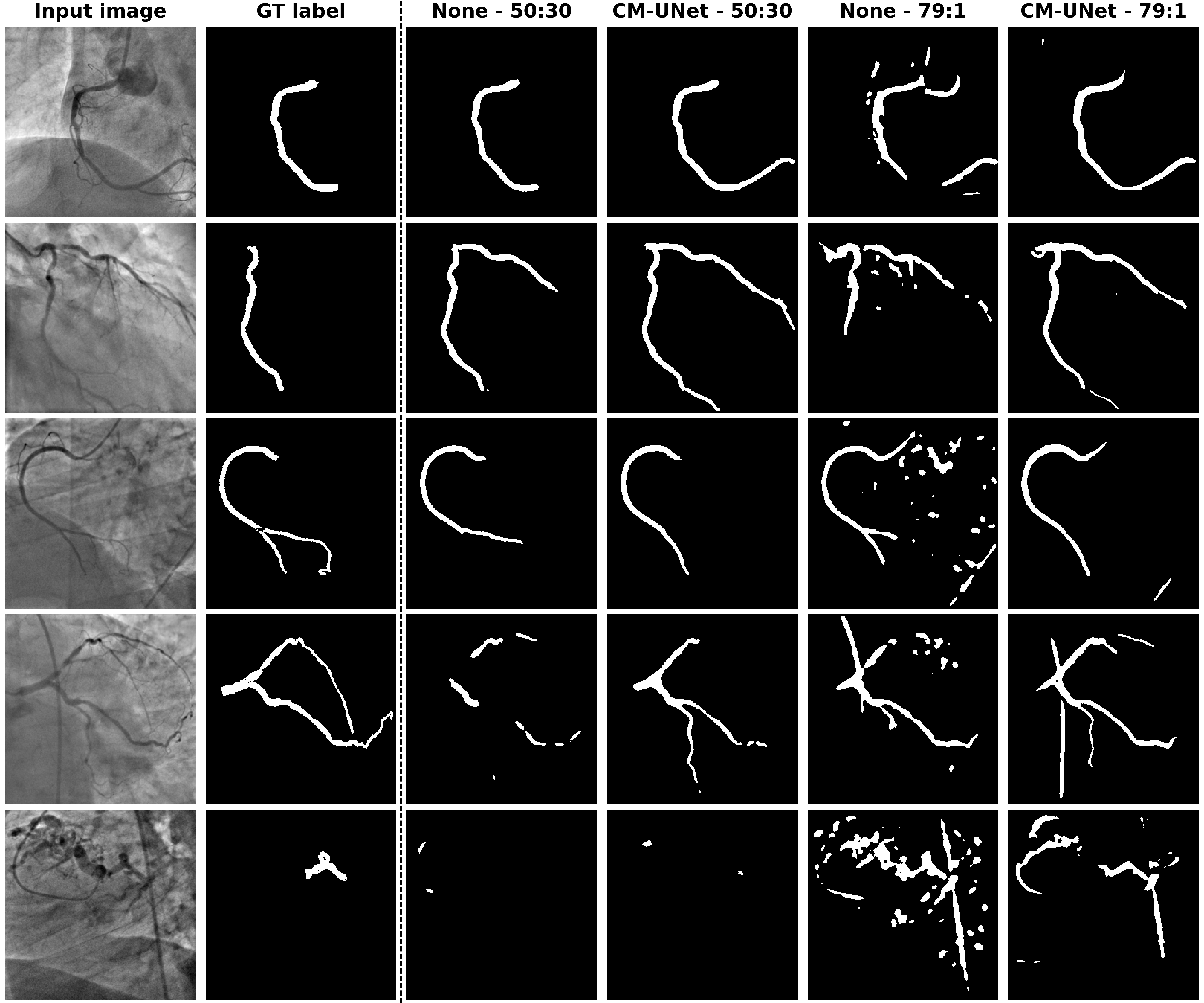}
    \caption{Segmentation examples of major vessels using UNet (without SSL) and CM-UNet, comparing the 50:30 and 79:1 pPT-FT ratios.}
    \label{fig:image}    
\end{figure}

Figure~\ref{fig:image} illustrates test set segmentation examples, including input ICA images, GT labels, and segmentation outputs from models with no pre-training (None) and CM-UNet, evaluated under different PT-FT ratios. For the 50:30 ratio, CM-UNet closely aligns with GT labels, accurately capturing vessel structures and maintaining continuity, whereas the non-pre-trained model produces fragmented and less precise outputs. Under the 79:1 ratio, where fine-tuning data is limited, CM-UNet continues to outperform, preserving overall vessel morphology despite minor discontinuities. 
In contrast, the non-pre-trained model exhibits severe degradation, with fragmented and incomplete segmentations. The last row shows the limits of CM-UNet under challenging conditions, such as poor contrast and faint vessel boundaries, where both models produce sparse or noisy segmentations.

\begin{figure}[h]
    \centering
    \includegraphics[width = \columnwidth]{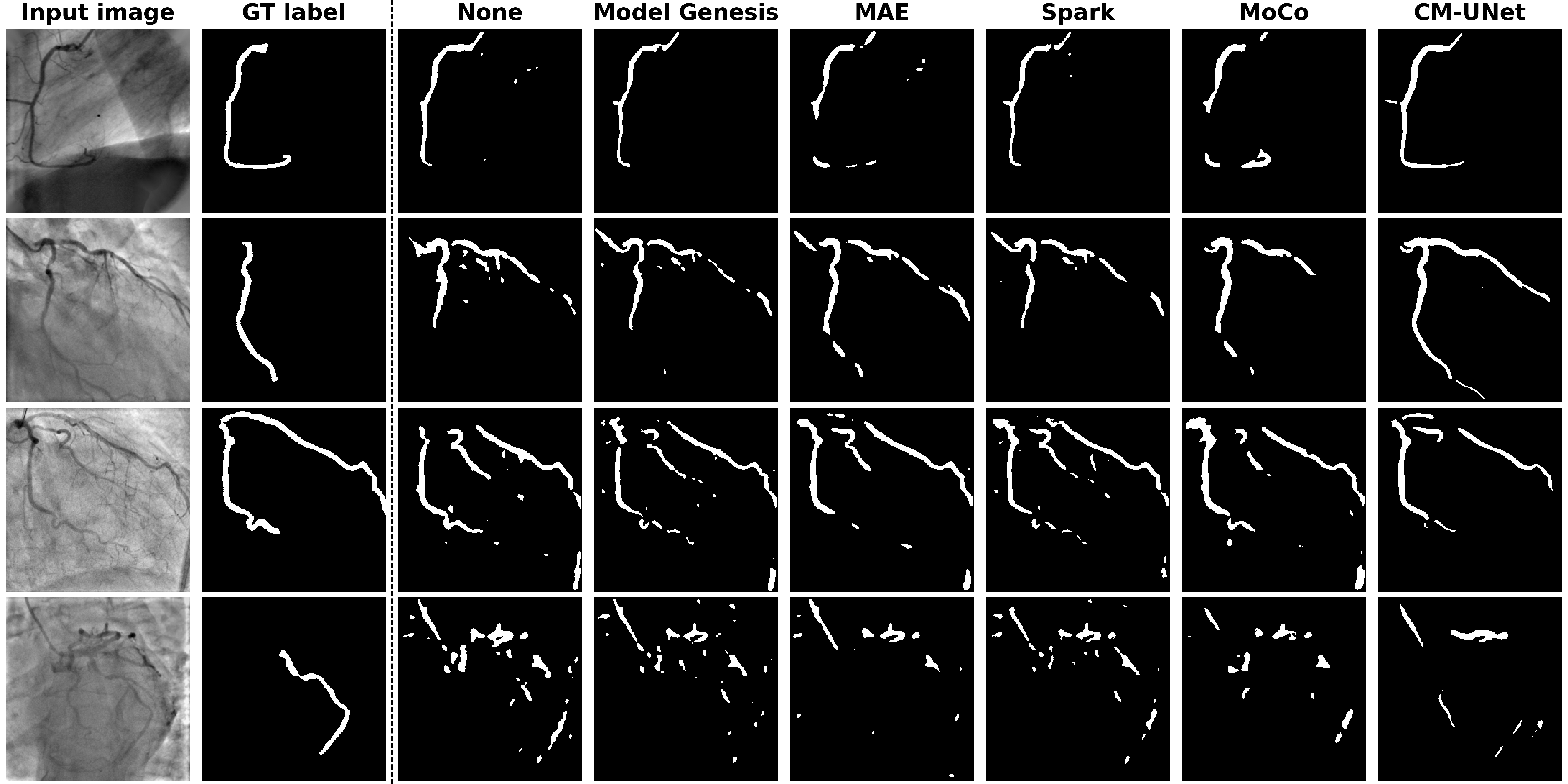}
    \caption{Segmentation examples of major vessels comparing UNet (without SSL) and all SSL methods using 79:1 PT-FT ratios.}
    \label{fig:visu_ssl}    
\end{figure}

Figure~\ref{fig:visu_ssl} visualizes the reconstruction results from various SSL models, including Model Genesis, MAE, SparK, and MoCo, to assess their effectiveness in learning robust representations. Among these, CM-UNet achieves the best performance, exhibiting the fewest discontinuities and the most accurate vessel reconstructions. MoCo and MAE follow as the next best methods, though they fall short of CM-UNet in preserving structural integrity. As seen in the last row, all models struggle with unclear vessel demarcation.

\section{Discussion and conclusion}

\label{sec:discussion}
In this work, we propose an SSL-based model, CM-UNet, to address the challenge of segmentation in medical imaging, particularly when annotated data is limited.
Our findings emphasize the critical role of data quantity and provide insights into selecting the most effective approach based on the available data. 
The proposed CM-UNet architecture exhibited significant enhancements in segmentation performance with a fine-tuning dataset of 500 images. Specifically, it achieved a 4.3\% increase in Dice coefficient when pre-trained on the FAME2 dataset, a statistically significant improvement (p-value $<$ 0.05) over the non-pre-trained model. Furthermore, fine-tuning the model with only 18 images instead of 500 led to a Dice coefficient decrease of only 15.2\%. This contrasts sharply with the baseline non-pre-trained models, which experienced a far greater drop of 46.5\%. This difference underscores the robustness of CM-UNet, even when fine-tuned with limited data, highlighting its ability to adapt and generalize effectively in settings with scarce labeled data.

CM-UNet shows strong potential for clinical translation. Requiring minimal annotations, it can quickly adapt across imaging protocols and machine-specific settings, making it well-suited for deployment in diverse clinical environments. CM-UNet could serve as a valuable tool in computer-assisted diagnosis of CAD, particularly in settings where expert resources are limited or rapid decision-making is required.

This work still has some limitations. Future work could extend the CM-UNet strategy to other datasets, medical imaging modalities, and segmentation tasks to assess its generalizability across diverse domains. An ablation study of the reconstruction and contrastive components could provide valuable insight into the contribution of each pre-training strategy. Additionally, incorporating post-processing techniques to address both discontinuity and over-segmentation of arteries segmentation simultaneously, could further improve segmentation performance, making it more robust and reliable for complex clinical scenarios.

In conclusion, our study highlights the effectiveness of CM-UNet in enhancing segmentation performance, particularly in data-scarce scenarios, by leveraging SSL. These findings demonstrate its robustness and adaptability, enabling a wider range of applications in medical imaging.

\addtolength{\textheight}{-12cm}   



\bibliographystyle{IEEEtran}
\bibliography{ref} 

\end{document}